\documentclass[12pt, centerh1]{article}
\usepackage{newtxtext}
\usepackage{amsmath,natbib,colonequals}
\usepackage{amsfonts,mathrsfs,moreverb,lipsum}
\usepackage{multimedia}
\usepackage{graphicx}

\newcommand{\btheta}{{\boldsymbol{\theta}}}
\newcommand{\bvtheta}{{\boldsymbol{\vartheta}}}
\newcommand{\pig}{\pi_g}

\newcommand{\bfx}{{\bf x}}

\newcommand{\vecmu}{\mbox{\boldmath$\mu$}}

\newcommand{\vecx}{\mathbf{x}}

\newcommand{\vecX}{\mathbf{X}}
\newcommand{\vecU}{\mathbf{U}}

\newcommand{\vecW}{\mathbf{W}}

\newcommand{\vecz}{\mathbf{z}}

\newcommand{\load}{\mathbf\Lambda}
\newcommand{\matsig}{\mathbf\Sigma}
\newcommand{\matPsi}{\mathbf\Psi}

\newcommand{\diag}{\,\mbox{diag}}
\newcommand{\tr}{\,\mbox{tr}}

\newcommand{\vecA}{\mathbf{A}}
\newcommand{\vecB}{\mathbf{B}}

\newcommand{\vecS}{\mathbf{S}}

\newcommand{\matm}{\mathbf{M}}

\newcommand{\ident}{\mathbf{I}}
\newcommand{\del}{\mathbf\Delta}
\newcommand{\vecepsilon}{\mbox{\boldmath$\varepsilon$}}

\newcommand{\vecOmega}{\mathbf\Omega}

\newcommand{\vecc}{\text{vec}}
\newcommand{\fX}{\mathscr{X}}
\newcommand{\fV}{\mathscr{V}}
\newcommand{\fU}{\mathscr{U}}
\newcommand{\fE}{\mathscr{E}}
\newcommand{\fY}{\mathscr{Y}}

\oddsidemargin=0mm\usepackage{natbib}

\title{A Mixture of Matrix Variate Bilinear Factor Analyzers}
\author{Michael P.B.\ Gallaugher and Paul D.\ McNicholas}
\date{\small Dept.\ of Mathematics \& Statistics, McMaster University, Hamilton, Ontario, Canada.}

\begin{document}

\maketitle
\begin{abstract}Over the years data has become increasingly higher dimensional, which has prompted an increased need for dimension reduction techniques. This is perhaps especially true for clustering (unsupervised classification) as well as semi-supervised and supervised classification. Although dimension reduction in the area of clustering for multivariate data has been quite thoroughly discussed within the literature, there is relatively little work in the area of three-way, or matrix variate, data.  Herein, we develop a mixture of matrix variate bilinear factor analyzers (MMVBFA) model for use in clustering high-dimensional matrix variate data. This work can be considered both the first matrix variate bilinear factor analysis model as well as the first MMVBFA model. Parameter estimation is discussed, and the MMVBFA model is illustrated using simulated and real data.\end{abstract}

\section{Introduction}
Dimensionality is an ever present concern with data becoming increasingly higher dimensional over the last few years. To combat this issue, dimension reduction techniques have become very important tools, especially in the area of clustering (unsupervised classification) as well as semi-supervised and supervised classification. For multivariate data, the mixture of factor analyzers model has proved to be very useful in this regard as the model performs clustering and dimension reduction simultaneously, details in Section~2. However, there is relative paucity in the area of dimension reduction for use in model-based clustering for matrix variate data. Matrix variate distributions have been shown to be useful for modelling three-way data such as images and multivariate longitudinal data; however, the methods presented in the literature suffer from dimensionality concerns. In this paper, we present a mixture of matrix variate bilinear factor analyzers (MMVBFA) model for use in clustering higher dimensional matrix variate data. The matrix variate bilinear factor analyzers model can be viewed as a generalization of bilinear principal component analysis \citep[BPCA;][]{zhao12}, and contains BPCA as a special case. An alternating expectation conditional maximization (AECM) algorithm \citep{meng97} is used for parameter estimation. The proposed method is illustrated using both simulated and real datasets.
\vskip14pt
\section{Background}
\subsection{Model-Based Clustering}
Model-based clustering makes use of a finite mixture model. A $G$-component finite mixture model assumes a random variate $\vecX$ has density
$$
f(\vecx~|~\bvtheta)=\sum_{g=1}^G\pig f_g(\vecx~|~\btheta_g),
$$
where $\bvtheta=\left(\pi_1,\pi_2,\ldots,\pi_G,\btheta_1,\btheta_2,\ldots,\btheta_G\right)$, $f_g(\cdot)$ is the $g$th component density, and $\pig>0$ is the $g$th mixing proportion such that $\sum_{i=1}^G\pig=1$.
The association between clustering and mixture models, as discussed in \cite{mcnicholas16a}, can be traced all the way back to \cite{tiedeman55}. The earliest use of a mixture model, specifically a Gaussian mixture model, for model-based clustering can be found in \cite{wolfe65}.  Other early work in this area can be found in \cite{baum70} and \cite{scott71} with a recent review given by \cite{mcnicholas16b}. 

The Gaussian mixture model is well-established for clustering both multivariate and matrix variate data because of its mathematical tractability; however, there are many examples of non-Gaussian distributions used for clustering. In the multivariate case, work has been done using symmetric component densities that parameterize concentration (tail weight), e.g., the $t$ distribution \citep{peel00,andrews11a,andrews12,steane12,morris13a,lin14} and the power exponential distribution \citep{dang15}. Furthermore, there has been some work in the area of multivariate skewed-distributions, for example the normal-inverse Gaussian distribution \citep{karlis09,subedi14,ohagan16}, the skew-$t$ distribution \citep{lin10,vrbik12,vrbik14,murray14b,murray14a,lee14,lee16,murray17a}, the shifted asymmetric Laplace distribution \citep{morris13b,franczak14}, the generalized hyperbolic distribution \citep{browne15,wei19}, the variance-gamma distribution \citep{smcnicholas17}, and the joint generalized hyperbolic distribution \citep{tang18}.

In the area of matrix variate data, \cite{viroli11} considers a mixture of matrix variate normal distributions for clustering, 
and \cite{dougru16} consider a mixture of matrix variate $t$ distributions. More recently, \cite{gallaugher17c} consider mixtures of four skewed matrix variate distributions, specifically the matrix variate skew-$t$ distribution \citep{gallaugher17a} as well as generalized hyperbolic, variance-gamma and normal-inverse Gaussian distributions \citep{gallaugher17b}, with an application in handwritten digit recognition. As pointed out by \cite{gallaugher17c}, these approaches are limited by the dimensionality of the data and the work described herein aims to help address that limitation.

\subsection{Matrix Variate Normal Distribution}\label{sec:22} 
An $n\times p$ random matrix $\fX$ follows a matrix variate normal distribution with location parameter $\matm$ and scale matrices $\del$ and $\vecOmega$ of dimensions $n\times n$ and $p\times p$, respectively, denoted by $\mathcal{N}_{n\times p}(\matm, \del, \vecOmega)$, if the density of $\fX$ can be written
\begin{equation*}
\varphi_{n\times p}(\vecX|\matm, \del, \vecOmega)=\frac{1}{(2\pi)^{\frac{np}{2}}|\del|^{\frac{p}{2}}|\vecOmega|^{\frac{n}{2}}}\exp\left\{-\frac{1}{2}\tr\left(\del^{-1}(\vecX-\matm)\vecOmega^{-1}(\vecX-\matm)'\right)\right\}.
\end{equation*}
One notable property of the matrix variate normal distribution \citep{harrar08} is
\begin{equation}
\fX\sim \mathcal{N}_{n\times p}(\matm,\del,\vecOmega) \iff \vecc(\fX)\sim \mathcal{N}_{np}(\vecc(\matm),\vecOmega\otimes \del),
\label{eq:normprop}
\end{equation}
where $\mathcal{N}_{np}(\cdot)$ is the multivariate normal density with dimension $np$, $\vecc(\cdot)$ is the vectorization operator, and $\otimes$ is the Kronecker product.

\subsection{Mixture of Factor Analyzers Model}
For the purpose of this subsection, we temporarily revert back to the notation where $\vecX_i$ represents a $p$-dimensional random vector, with $\bfx_i$ as its realization. The factor analysis model for $\vecX_1,\ldots,\vecX_n$ is given by
$$
\vecX_i=\vecmu+\load\vecU_i+\vecepsilon_i,
$$
where $\vecmu$ is a location vector, $\load$ is a $p\times q$ matrix of factor loadings with $q<p$, $\vecU_i\sim\mathcal{N}_q({\bf 0},\ident)$ denotes the latent factors, $\vecepsilon_i\sim \mathcal{N}_q({\bf 0},\matPsi)$, where $\matPsi=\text{diag}(\psi_1,\psi_2,\ldots,\psi_p)$, and the $\vecU_i$ and the $\vecepsilon_i$ are each independently distributed and are independent of one another. Under this model, the marginal distribution of $\vecX_i$ is $\mathcal{N}_p(\vecmu,\load\load'+\matPsi)$. Probabilistic principal component analysis (PPCA) arises as a special case with the isotropic constraint $\matPsi=\psi\ident$, $\psi\in\mathbb{R}^+$ \citep{tipping99a}. 

\cite{ghahramani97} develop the mixture of factor analyzers model, which is a Gaussian mixture model with covariance structure $\matsig_g=\load_g\load_g'+\matPsi$. \cite{mclachlan00a} utilize the more general structure $\matsig_g=\load_g\load_g'+\matPsi_g$. \cite{tipping99b} introduce the closely-related mixture of PPCAs with $\matsig_g=\load_g\load_g'+\psi_g\ident$. \cite{mcnicholas08} construct a family of eight parsimonious Gaussian models by considering the constraint $\load_g=\load$ in addition to $\matPsi_g=\matPsi$ and $\matPsi_g=\psi_g\ident$. There has also been work on extending the mixture of factor analyzers to other distributions, such as the skew-$t$ distribution \citep{murray14b,murray17a}, the generalized hyperbolic distribution \citep{tortora16}, the skew-normal distribution \citep{lin16}, the variance-gamma distribution \citep{smcnicholas17}, and others \citep[e.g.,][]{murray17a}.

\subsection{Previous Work on Matrix Variate Factor Analysis}
\cite{xie08} and \cite{yu08} consider a matrix variate extension of PPCA in a linear fashion. For $N$ independent $n\times p$ random matrices $\fX_1,\ldots,\fX_N$, the model assumes
\begin{equation}
\fX_i=\matm+\vecA\fU_i\vecB'+\fE_i,
\label{eq:MVPPCA}
\end{equation}
where $\matm$ is an $n\times p$ location matrix, $\vecA$ is an $n\times q$ matrix of column factor loadings, $\vecB$ is a $p\times r$ matrix of row factor loadings, $\fU_i\sim \mathcal{N}_{q\times r}({\bf 0},\ident_q,\ident_r)$, and $\fE_i\sim\mathcal{N}_{n\times p}({\bf 0},\sigma\ident_n,\sigma\ident_p)$, with $\sigma\in\mathbb{R}^+$. Note that the $\fU_i$ and the $\fE_i$ are each independently distributed and are independent of one another. The main disadvantage of this model is that, in general, $\fX_i$ does not follow a matrix variate normal distribution. 

\cite{zhao12} present bilinear probabilistic principal component analysis (BPPCA) which extends \eqref{eq:MVPPCA} by adding two projected error terms. The resulting model assumes 
\begin{equation}
\fX_i=\matm+\vecA\fU_i\vecB'+\vecA\fE_i^B+\fE_i^A\vecB'+\fE_i,
\label{eq:BFA}
\end{equation}
where 
$\fE_i^B\sim \mathcal{N}_{q\times p}({\bf 0},\ident_q,\sigma_B\ident_p)$, $\fE_i^A\sim \mathcal{N}_{n\times r}({\bf 0},\sigma_A\ident_n,\ident_r)$, $\fE_i\sim \mathcal{N}_{n\times p}(0,\sigma_{A}\ident_n,\sigma_B\ident_{p})$, with $\sigma_A\in\mathbb{R}^+$ and $\sigma_B\in\mathbb{R}^+$, and the other terms are as defined for \eqref{eq:MVPPCA}. In this model, each of the $\fU_i$, $\fE_i^B$, $\fE_i^A$ and $\fE_i$ are independently distributed and all are independent of each other.
It is important to note that the term ``column factors'' refers to reduction in the dimension of the columns, which is equivalent to the number of rows, and not a reduction in the number of columns. Likewise, the term ``row factors'' refers to the reduction in the dimension of the rows (i.e., in the number of columns).
As discussed by \cite{zhao12}, the interpretation of the $\fE_i^B$ and $\fE_i^A$ are the row and column noise, respectively, whereas $\fE_i$ is the common noise. It can be shown using property \eqref{eq:normprop} that under this model $\fX_i\sim\mathcal{N}_{n\times p}(\matm,\vecA\vecA'+\sigma_A\ident_n,\vecB\vecB'+\sigma_B\ident_p)$. Note that the covariance structure for the two covariance matrices of this matrix variate normal distribution are analogous to the covariance structure for the (multivariate) factor analysis model.

\section{Methodology}
\subsection{MMVBFA Model}
A MMVBFA model is derived here by extending \eqref{eq:BFA}. Specifically, we remove the isotropic constraint and assume that
\begin{equation}
\fX_i=\matm_g+\vecA_g\fU_{ig}\vecB_g'+\vecA_g\fE_{ig}^B+\fE^A_{ig}\vecB_g'+\fE_{ig}
\label{eq:FAmod}
\end{equation}
with probability $\pi_g$, for $g=1,\ldots,G$, where $\matm_g$ is an $n\times p$ location matrix, $\vecA_g$ is an $n\times q$ column factor loading matrix, with $q<n$, $\vecB_g$ is a $p\times r$ row factor loading matrix, with $r<p$, and
\begin{equation*}\begin{split}
\fU_{ig}&\sim \mathcal{N}_{q\times r}({\bf 0},\ident_q,\ident_r),\\
\fE_{ig}^{B}&\sim  \mathcal{N}_{q\times p}({\bf 0},\ident_q,\matPsi_g),\\
\fE_{ig}^{A}&\sim  \mathcal{N}_{n\times r}({\bf 0},\matsig_g,\ident_r),\\
\fE_{ig}&\sim   \mathcal{N}_{n\times p}({\bf 0},\matsig_g,\matPsi_g),
\end{split}\end{equation*}
each independently distributed and independent of each other, $\matsig_g=\diag\{\sigma_{1g},\ldots,\sigma_{ng}\}$, with $\sigma_{ig}\in\mathbb{R}^+$, and $\matPsi_g=\diag\{\psi_{1g}, \ldots, \psi_{pg}\}$, with $\psi_{ig}\in\mathbb{R}^+$.

Let $\vecz_i=(z_{i1},\ldots,z_{iG})'$ denote the component membership for $\vecX_i$, where
$$ 
z_{ig}=
\begin{cases}
1 & \mbox{if } \vecX_i \mbox{ belongs to component } g,\\
0 &\mbox{otherwise},
\end{cases}
$$ for $i=1,\ldots,N$ and $g=1,\ldots,G$. Using the vectorization of $\fX_i$, and property \eqref{eq:normprop}, it can be shown that 
$$
\fX_i|z_{ig}=1 \sim \mathcal{N}_{n\times p}(\matm_g,\matsig_g+\vecA_g\vecA_g',\matPsi_g+\vecB_g\vecB_g').
$$
Therefore, the density of $\fX_i$ can be written
$$
f(\vecX_i|\bvtheta)=\sum_{g=1}^G\pig \varphi_{n\times p}(\vecX_i|\matm_g,\matsig_g+\vecA_g\vecA_g',\matPsi_g+\vecB_g\vecB_g'),
$$
where $\varphi_{n\times p}(\cdot)$ denotes the $n\times p$ matrix variate normal density (see Section~\ref{sec:22}).
Following a similar procedure to that described by \cite{zhao12}, by introducing latent matrix variables $\fY_{ig}^B$ and $\fV_{ig}^B$, \eqref{eq:FAmod} can be written 
\begin{equation*}\begin{split}
\fX_i&=\matm_g+\vecA_g\fY_{ig}^B+\fV_{ig}^B,\\
\fY_{ig}^B&=\fU_{ig}\vecB_g'+\fE_{ig}^B,\\
\fV_{ig}^B&=\fE_{ig}^A\vecB_{g}'+\fE_{ig}.
\end{split}\end{equation*}
The two-stage interpretation of this formulation of the model is the same as that given by \cite{zhao12}, where this can viewed as first projecting $\fX_i$ in the column direction onto the latent matrix $\fY_{ig}^B$, and then $\fY_{ig}^B$ and $\fV_{ig}^B$ are further projected in the row direction.
Likewise, introducing $\fY_{ig}^A$ and $\fV_{ig}^A$, \eqref{eq:FAmod} can be written
\begin{equation*}\begin{split}
\fX_i&=\matm_g+\fY_{ig}^A\vecB_g'+\fV_{ig}^A,\\
\fY_{ig}^A&=\vecA_g\fU_{ig}+\fE_{ig}^A,\\
\fV_{ig}^A&=\vecA_g\fE_{ig}^B+\fE_{ig}.
\end{split}\end{equation*}
The interpretation is the same as before only we project in the row direction first followed by the column direction.
It can be shown that
$$
\fY_{ig}^B|\vecX_i,z_{ig}=1\sim\mathcal{N}_{q\times p}({\vecW^{A}_g}^{-1}\vecA_g'\matsig_g^{-1}(\vecX_i-\matm_g),{\vecW^{A}_g}^{-1},\load_{\vecB_g})
$$
and
$$
\fY_{ig}^A|\vecX_i,z_{ig}=1\sim\mathcal{N}_{n\times r}((\vecX_i-\matm_g)\matPsi_g^{-1}\vecB_g{\vecW_g^{B}}^{-1},\load_{\vecA_g},{\vecW^{B}_g}^{-1}),
$$
where $\vecW^{A}_g=\ident_{q}+\vecA_g'\matsig_g^{-1}\vecA_g$, $\vecW^{B}_g=\ident_r+\vecB_{g}'\matPsi_g^{-1}\vecB_g$, $\load_{\vecA_g}=\matsig_g+\vecA_g\vecA_g'$, and $\load_{\vecB_g}=\matPsi_g+\vecB_g\vecB_g'$

\subsection{Parameter Estimation}
Suppose we observe $N$ observations $\vecX_1, \vecX_2, \ldots, \vecX_N$ then the log-likelihood is given by
\begin{equation}
\mathcal{L}(\bvtheta)=\sum_{i=1}^N\log\sum_{g=1}^G \pig\varphi_{n\times p}(\vecX_i|\matm_g,\matsig_g+\vecA_g\vecA_g',\matPsi_g+\vecB_g\vecB_g').
\label{eq:inclik}
\end{equation}
To maximize \eqref{eq:inclik}, the observed data is viewed as incomplete and an AECM is then to maximize \eqref{eq:inclik}. There are three different sources of missingingness: the component memberships $\vecz_1,\ldots,\vecz_n$ as well as the latent matrix variables $\fY_{ig}^B$ and $\fY_{ig}^A$. A three-stage AECM algorithm is now described for parameter estimation.\\

\noindent \textbf{AECM Stage 1}: In the first stage, the complete-data is taken to be the observed matrices $\vecX_1,\ldots,\vecX_N$ and the component memberships $\vecz_1,\ldots,\vecz_N$, and the updates for $\pi_g$ and $\matm_g$ are calculated. The complete-data log-likelihood in the first stage is
$$
\ell_{1}=C_1+\sum_{g=1}^G\sum_{i=1}^N z_{ig}\left\{\log\pig-\frac{1}{2} \tr[\load_{\vecA_g}^{-1}(\vecX_i-\matm_g)\load_{\vecB_g}^{-1}(\vecX_i-\matm_g)']\right\},
$$
where $C_1$ is a constant with respect to $\pi_g$ and $\matm_g$.
In the E-Step, the updates for the component memberships $z_{ig}$ are given by
$$
\hat{z}_{ig}=\frac{\pig\varphi_{n\times p}(\vecX_i~|~{\matm}_g,{\load}_{\vecA_g},{\load}_{\vecB_g})}{\sum_{h=1}^G\pig\varphi_{n\times p}(\vecX_i~|~{\matm}_h,{\load}_{\vecA_h},{\load}_{\vecB_h})},
$$
where $\varphi_{n\times p}(\cdot)$ denotes the $n\times p$ matrix variate normal density.
As usual, these expectations $\hat{z}_{ig}$ are calculated using the current estimates of the parameters.
In the CM-step, the updates for $\pi_g$ and $\matm_g$ are calculated using
$$
\hat{\pi}_g=\frac{N_g}{N} \quad\text{ and }\quad\hat{\matm}_g=\frac{1}{N_g}\sum_{i=1}^N\hat{z}_{ig}\vecX_{i},
$$
respectively, where $N_g=\sum_{i=1}^N\hat{z}_{ig}$.\\

\noindent \textbf{AECM Stage 2}: In the second stage, the complete-data is taken to be the observed $\vecX_1,\ldots,\vecX_N$, the component memberships $\vecz_1,\ldots,\vecz_N$ and the $n\times q$ latent matrices $\fY_{ig}^B$. The complete-data log-likelihood is then
\begin{equation*}\begin{split}
\ell_{2}=&C_2-\frac{N_gp}{2}\log|\matsig_g|-\frac{1}{2}\sum_{g=1}^G\sum_{i=1}^Nz_{ig}\text{tr}\big[\matsig_g^{-1}(\vecX_i-\matm_g)\load_{\vecB_g}^{-1}(\vecX_i-\matm_g)'\\
&\ -\matsig_g^{-1}\vecA_g\fY_{ig}^B\load_{\vecB_g}^{-1}(\vecX_i-\matm_g)'-\matsig_{g}^{-1}(\vecX_i-\matm_g)\load_{\vecB_g}^{-1}{\fY_{ig}^B}'\vecA_g'\\&+\matsig_g^{-1}\vecA_g\fY_{ig}^B\load_{\vecB_g}^{-1}{\fY_{ig}^{B}}'\vecA_g'\big],
\end{split}\end{equation*}
where $C_2$ is constant with respect to the parameters being updated.
In the E-Step, the following expectations are calculated:
\begin{equation*}\begin{split}
a^B_{ig}&\colonequals\mathbb{E}[\fY_{ig}^B~|~\vecX_{i},z_{ig}=1]={\vecW^{A}_g}^{-1}\vecA_g'\matsig_g^{-1}(\vecX_i-\matm_g),\\
b^B_{ig}&\colonequals\mathbb{E}[\fY_{ig}^B\hat{\load}_{\vecB_g}^{-1}{\fY_{ig}^B}'~|~\vecX_{i},z_{ig}=1]=p{\vecW^{A}_g}^{-1}+a^B_{ig}\load_{\vecB_g}^{-1}{a^B_{ig}}'.
\end{split}\end{equation*}
As usual, these expectations are calculated using the current estimates of the parameters.
In the CM-step, $\vecA_g$ and $\matsig_g$ are updated via
\begin{equation*}\begin{split}
\hat{\vecA}_g&=\sum_{i=1}^N\hat{z}_{ig}(\vecX_i-\hat{\matm}_g)\hat{\load}_{\vecB_g}^{-1}{a_{ig}^B}'\left(\sum_{i=1}^N \hat{z}_{ig}b_{ig}^B\right)^{-1}\quad\text{and}\quad
\hat{\matsig}_g=\frac{1}{N_gp}\diag\{\hat{\vecS}^B_g\},
\end{split}\end{equation*}
respectively, where 
$$
\vecS^B_g=\sum_{i=1}^N\hat{z}_{ig}\big[(\vecX_i-\hat{\matm}_g)\hat{\load}_{\vecB_g}^{-1}(\vecX_i-\hat{\matm}_g)'-\hat{\vecA}_ga_{ig}^B\hat{\load}_{\vecB_g}^{-1}(\vecX_i-\hat{\matm_g})'\big].
$$

\noindent \textbf{AECM Stage 3}: In the last stage of the AECM algorithm, the complete-data is taken to be the observed $\vecX_1,\ldots,\vecX_N$, the component memberships $\vecz_1,\ldots,\vecz_N$ and the $p\times r$ latent matrices $\fY_{ig}^A$. In this step, the complete-data log-likelihood is
\begin{equation*}\begin{split}
\ell_{3}=&C_3-\frac{N_g n}{2}\log|\matPsi_g|-\frac{1}{2}\sum_{g=1}^G\sum_{i=1}^N z_{ig}\tr\big[\matPsi_g^{-1}(\vecX_i-\matm_g)'\load_{\vecA_g}^{-1}(\vecX_i-\matm_g)\\
&-\matPsi_g^{-1}\vecB_g{\fY_{ig}^A}'\load_{\vecA_g}^{-1}(\vecX_i-\matm_g)-\matPsi_g^{-1}(\vecX_i-\matm_g)'\load_{\vecA_g}^{-1}{\fY_{ig}^A}\vecB_g'\\&+\matPsi_g^{-1}\vecB_g{\fY_{ig}^A}'\load_{\vecA_g}^{-1}{\fY_{ig}^{A}}\vecB_g'\big],
\end{split}\end{equation*}
where $C_3$ is constant with respect to the parameters being updated.
In the E-Step, expectations similar to those in the second step are calculated, i.e.,
$$
a^A_{ig}:=\mathbb{E}[\fY_{ig}^A~|~\vecX_{i},z_{ig}=1]=(\vecX_i-\matm_g)\matPsi_g^{-1}\vecB_g{\vecW_g^{B}}^{-1}
$$
and
$$
b^A_{ig}:=\mathbb{E}[{\fY_{ig}^A}'\load_{\vecB_g}^{-1}\fY_{ig}^A~|~\vecX_{i},z_{ig}=1]=n{\vecW^{B}_g}^{-1}+{a^A_{ig}}'\load_{\vecA_g}^{-1}{a^A_{ig}}.
$$
As usual, these expectations are calculated using the current estimates of the parameters.
In the CM-step, we update $\vecB_g$ and $\matPsi_g$ by
\begin{equation*}\begin{split}
\hat{\vecB}_g&=\sum_{i=1}^N\hat{z}_{ig}(\vecX_i-\hat{\matm}_g)'\hat{\load}_{\vecA_g}^{-1}{a_{ig}^A}\left(\sum_{i=1}^N \hat{z}_{ig}b_{ig}^A\right)^{-1}\quad\text{and}\quad
\hat{\matPsi}_g=\frac{1}{N_gn}\diag\{\hat{\vecS}^A_g\},
\end{split}\end{equation*}
respectively, where 
$$
\vecS^A_g=\sum_{i=1}^N\hat{z}_{ig}\big[(\vecX_i-\hat{\matm}_g)'\hat{\load}_{\vecA_g}^{-1}(\vecX_i-\hat{\matm}_g)-\hat{\vecB}_g{a_{ig}^A}'\hat{\load}_{\vecA_g}^{-1}(\vecX_i-\hat{\matm_g})\big].
$$

\subsection{Semi-Supervised Classification}
The MMVBFA model presented herein for clustering may also be used for semi-supervised classification. Suppose $N$ matrices are observed, and $K$ of these observations have known labels from one of $G$ classes. Following \cite{mcnicholas10}, without loss of generality, order the matrices so that the first $K$ have known labels and the remaining observations have unknown labels. The observed likelihood is then
\begin{equation*}\begin{split}
L(\bvtheta)=\prod_{i=1}^K\prod_{g=1}^G&\big[\pig\varphi_{n\times p}(\vecX_i|\matm_g,\matsig_g+\vecA_g\vecA_g',\matPsi_g+\vecB_g\vecB_g')\big]^{z_{ig}}\\
&\times \prod_{j=K+1}^N\sum_{h=1}^H\pi_h\varphi_{n\times p}(\vecX_j|\matm_h,\matsig_h+\vecA_h\vecA_h',\matPsi_h+\vecB_h\vecB_h').
\end{split}\end{equation*}
It is possible for $H\ne G$; however, for our analyses we assume that $H=G$. Parameter estimation then proceeds in a similar manner for the clustering scenario. For more information on semi-supervised classification refer to \cite{mcnicholas16a}.

\subsection{Model Selection, Initialization and Convergence}
For a typical dataset the number of components and/or the number of factors will not be known {a priori} and, therefore, we will have to select them. One common selection criterion is the Bayesian information criterion \cite[BIC;][]{schwarz78} and is given by
$$
\text{BIC}=2\ell(\hat{\bvtheta})-\rho\log N,
$$
where $\ell(\hat{\bvtheta})$ is the maximized log-likelihood and $\rho$ is the number of free parameters. The BIC is used as the selection criterion for all of our analyses. 

To initialize the AECM algorithm, we employ an alternating emEM strategy \citep{biernacki03}. This consists of running the AECM algorithm for a small number of iterations for different random starting values of the parameters and then using the parameters that maximize the likelihood to continue with the AECM algorithm until convergence.

The simplest convergence criterion would be to use lack of progress in the log-likelihood, however; it is possible for the log-likelihood to ``plateau'' and then increase again thus terminating the algorithm prematurely \citep[see][]{mcnicholas16a}. One alternative is to use a criterion based on the Aitken acceleration \citep{aitken26}. The acceleration at iteration $t$ is
$$
a^{(t)}=\frac{l^{(t+1)}-l^{(t)}}{l^{(t)}-l^{(t-1)}},
$$
where $l^{(t)}$ is the observed likelihood at iteration $t$. Now,
$$
l_{\infty}^{(t+1)}=l^{(t)}+\frac{(l^{(t+1)}-l^{(t)})}{1-a^{(t)}},
$$
is an estimate of the observed log-likelihood after many iterations, at iteration $t+1$ \citep[see][]{bohning94,lindsay95}. As in \cite{mcnicholas10a}, we terminate the algorithm when $l_{\infty}^{(k+1)}-l^{(k)}\in(0,\epsilon)$.
It is important to note that, in each AECM algorithm run for the analyses herein, we make the choice of $\epsilon$ based on the magnitude of the log-likelihood. Specifically, after running the 10 iterations of the emEM algorithm, we choose~$\epsilon$ to be four orders of magnitude lower than the log-likelihood. 

\subsection{Reduction in Number of Free Covariance Parameters}
Because the covariance structure of both covariance matrices in the MVVBFA model is equivalent to the covariance structure in the (multivariate) mixture of factor analyzers model, many of the results on the number of free covariance parameters may be used here.  Specifically there are $nq+n-q(q-1)/2$ free covariance parameters in $\load_{\vecA_g}$ and $pr+p-r(r-1)/2$ free covariance parameters in $\load_{\vecB_g}$ \citep{lawley62}.  Therefore, reduction in the number of free covariance parameters for the row covariance matrix is
$$
\frac{1}{2}n(n+1)-nq-n+\frac{1}{2}q(q-1)=\frac{1}{2}\big[(n-q)^2-(n+q)\big],
$$
which is positive for $(n-q)^2>n+q$.
Likewise for the column covariance matrix the reduction in the number of parameters is
$$
\frac{1}{2}p(p+1)-pr-p+\frac{1}{2}r(r-1)=\frac{1}{2}\big[(p-r)^2-(p+r)\big],
$$
which is positive for $(p-r)^2>p+r$.
In applications herein, the model is fit for a range of row factors and column factors. If the number of row or column factors chosen by the BIC is the maximum in that range, the relevant number of factors will be increased so long as the aforementioned conditions are met.

\section{Data Analyses}
\subsection{Simulations}
\subsubsection*{Simulation 1}
In the first simulation, $G=2$ groups are considered with $10\times7$ matrices. The mixing proportions are taken to be $\pi_1=\pi_2=0.5$, and we set $N\in\{200,400,800\}$. Observations are simulated from \eqref{eq:FAmod} with $q=2$ column factors and $r=3$ row factors. For each value of $N$, 50 datasets are simulated. For each dataset, for each $N$, the correct number of groups, column and row factors are selected. In addition, perfect classification is achieved ($\text{ARI}=1$). Note that the adjusted Rand index \citep[ARI;][]{rand71,hubert85} is often used to asses agreement between true and predicted classes; it takes a value of 1 for perfect class agreement and has expected value $0$ under random class assignment.  In Table~ \ref{tab:sim1}, we show the average value of $\|\matm_g-\hat{\matm}_g\|_1$, for $g=1,2$ and for each value of $N$, over the 50 datasets. Note that if $\vecW$ is an $n\times p$ matrix then 
$$
\|\vecW\|_1=\max_{1\le j\le p}\sum_{i=1}^n|w_{ij}|.
$$
As expected, the estimates of $\matm_g$ get closer to the true values as the sample size $N$ in increased. Moreover, the variability of $\|\matm_g-\hat{\matm}_g\|_1$  decreases as the sample size increases.
\begin{table}[!htb]
\centering
\caption{Average $\|\matm_g-\hat{\matm}_g\|_1$ values over 50 datasets, for $g=1,2$ and $N=200,400,800$, in Simulation 1, with standard deviations in parentheses.}
\begin{tabular}{lrrr}
\hline
&\multicolumn{3}{c}{$N$}\\
\cline{2-4}
$g$&200&400&800\\
\hline
1&13.97(3.61)&9.66(2.65)&6.48(1.69)\\
2&12.08(3.25)&7.45(1.79)&5.69(1.32)\\
\hline
\end{tabular}
\label{tab:sim1}
\end{table}

\subsubsection*{Simulation 2}
The second simulation considers $G=3$ groups with $28\times 17$ matrices. The mixing proportions are $\pi_1=\pi_3=0.4$ and $\pi_2=0.2$, and $N\in\{250,500,1000\}$. Again, 50 datasets are simulated for each $N$ with $q=2$ column factors and $r=3$ row factors. As in Simulation 1, the correct number of groups, column and row factors are chosen and perfect classification is achieved. In Table \ref{tab:sim2}, we again show the average 1-norms for the differences between the true and estimated location parameters.
\begin{table}[!htb]
\centering
\caption{Average $\|\matm_g-\hat{\matm}_g\|_1$ values over 50 datasets, for $g=1,2,3$ and $N=250,500,1000$, in Simulation 2, with standard deviations in parentheses.}
\begin{tabular}{lrrr}
\hline
&\multicolumn{3}{c}{$N$}\\
\cline{2-4}
$g$&250&500&1000\\
\hline
1&36.28(7.95)&26.36(5.12)&19.37(4.62)\\
2&55.23(11.75)&40.42(9.64)&29.30(6.26)\\
3&39.10(8.89)&27.09(6.37)&19.99(4.45)\\
\hline
\end{tabular}
\label{tab:sim2}
\end{table}

 \subsection{MNIST Digit Recognition}\label{sec:mnist}
We consider the $28 \times 28$ MNIST digit dataset \citep{lecun98}, which contains over 60,000 greyscale images of handwritten Arabic digits 0 to 9. The images are represented by $28 \times 28$ pixel matrices with greyscale intensities ranging from 0 to 255. Because of the lack of variability in the outer rows and columns, some random noise is added while adding 50 to each of the non-zero elements to avoid confusing the noise with a true signal.  
We are interested in comparing digit 1 to digit 7, as was considered in \cite{gallaugher17c}. Similar to \cite{gallaugher17c}, we consider semi-supervised classification with 25\%, 50\% and 75\% supervision. In each case, 25 datasets are considered, each consisting of 200 observations from each digit, and we fit the model for 10 to 20 column and row factors.

In Table \ref{tab:class}, we show an aggregated classification table between the true and predicted classifications at each level of supervision for the points considered unlabelled. As expected, slightly better classification performance is obtained when the level of supervision is increased. Moreover, there is a more substantial difference when going from 25\% supervision to 50\% supervision than from 50\% to 75\%. 
\begin{table}[!htb]
\centering
\caption{Cross-tabulations of true (1,7) versus predicted (P1, P7) classifications for the observations considered unlabelled in the MNIST data at each level of supervision, aggregated over all runs.}
\begin{tabular}{rcc|cc|cc}
\hline
&\multicolumn{2}{c|}{25\% Supervision}&\multicolumn{2}{c|}{50\% Supervision}&\multicolumn{2}{|c}{75\% Supervision}\\
\hline
&P1&P7& P1&P7&P1&P7\\
\hline
1&3550 & 173& 2449 &  53  &1232&26\\ 
  7 &200 & 3577&  51 & 2447 & 18&1221\\ 
\hline
\end{tabular}
\label{tab:class}
\end{table} 

Table~\ref{tab:ARI} shows the average ARI and misclassification rate (MCR) over the 25 datasets, with the respective standard deviations, for each level of supervision. We note that we obtain better results than \cite{gallaugher17c} even with a lower level of supervision; however, the results in \cite{gallaugher17c} were based on resized images due to dimensionality constraints whereas this analysis was performed on the original images.
\begin{table}[!ht]
\centering
\caption{Average ARI values and misclassification rates (MCR), with associated standard deviations in parentheses, for each level of supervision for the points considered unlabelled for the MNIST data, aggregated over all runs.}
\begin{tabular}{lrr}
\hline
 & $\overline{\text{ARI}}$ (std.\ dev.) &$\overline{\text{MCR}}$(std.\ dev.)\\ 
\hline
25\%&0.82(0.15)&0.050(0.046)\\
50\% & 0.92 (0.056) & 0.021 (0.015)\\
75\% & 0.93 (0.056) & 0.018 (0.015)\\
\hline
\end{tabular}
\label{tab:ARI}
\end{table}

In Table \ref{tab:factors} the frequency of the number of factors chosen for each level of supervision over the 25 datasets is shown. For the majority of the datasets, the number of row and column factors lie between 13 and 15.
\begin{table}[!htb]
\centering
\caption{Numbers of row and columns factors chosen for the MNIST dataset for 25\%,  50\% and 75\% supervision.}
\begin{tabular}{lccccccccccc}
\hline
&10&11&12&13&14&15&16&17&18&19&20\\
\hline
&\multicolumn{11}{c}{25\% Supervision}\\
\cline{2-12}
Row Factors &0&0&0&2&  7&  6&  4&  3&  2&  1&0\\
Column Factors &  0&0&2&  6&  7&  6&  3&  1& 0&0&0\\
\cline{2-12}
&\multicolumn{11}{c}{50\% Supervision}\\
\cline{2-12}
Row Factors &0&0&0&4&6&10&2&0&1&1&1\\ 
Column Factors & 0&0&2&9&7&5&1&1&0&0&0\\
\cline{2-12}
&\multicolumn{11}{c}{75\% Supervision}\\
\cline{2-12}
Row Factors &0&0&0&1&9&9&3&3&0&0&0\\
Column Factors &0&0&0&9&11&4&0&0&0&0&1\\
\hline
\end{tabular}
\label{tab:factors}
\end{table}

\begin{figure}[!htb]
\centering
\includegraphics[width=0.5\textwidth]{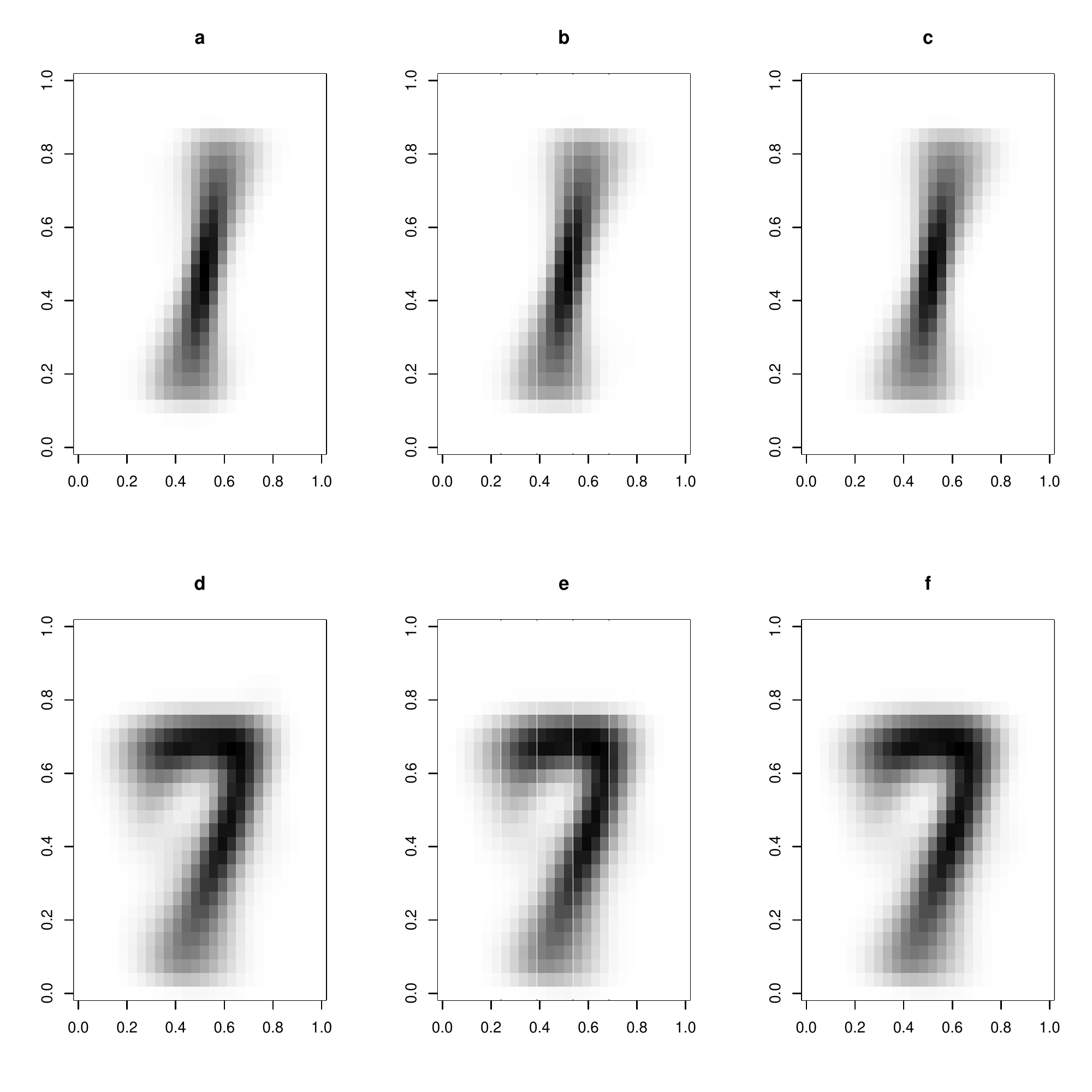}
\vspace{-0.2in}
\caption{Heatmaps for the average estimated location matrices taken over the 25 runs for digit 1 at 25\%, 50\% and 75\% supervision, respectively (a, b, c), and digit 7 at 25\%, 50\% and 75\% supervision, respectively (d, e, f).}
\label{fig:MNIST}
\end{figure}

Finally, in Figure \ref{fig:MNIST},  heatmaps are displayed for the average estimates of the location matrices over the 25 runs for each level of supervision for both digits. We see a slight increase in quality when going from 25\%  to 50\% supervision for digit 7 with the centre of the digit being a little smoother with 50\% supervision. There is no noticeable difference when going from 50\% to 75\% supervision. This similarity across the three levels of supervision illustrates the power of semi-supervised classification.

\subsection{Olivetti Faces Dataset}
\begin{figure}[!htb]
\centering
\includegraphics[width=1\textwidth]{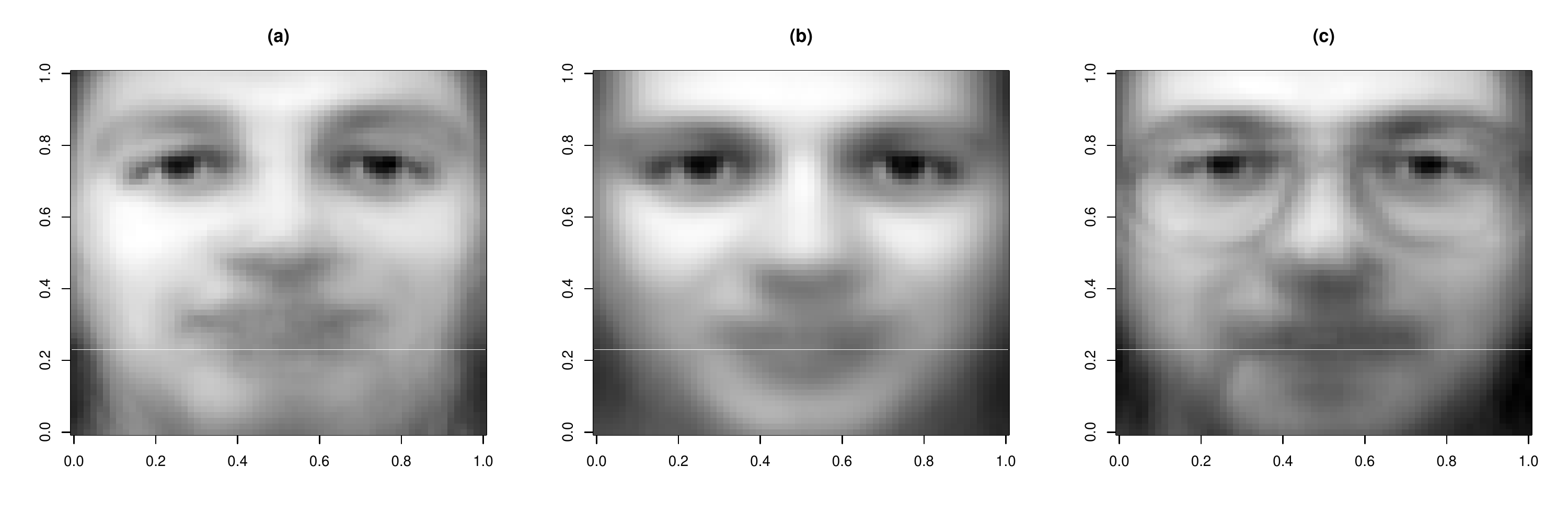}
\vspace{-0.4in}
\caption{Estimated location matrices for (a) component 1, (b) component 2, and (c) component 3 for the faces dataset.}
\label{fig:Face}
\end{figure}

Finally, consider the Olivetti faces dataset from the {\sf R} package {\tt RnavGraphImageData} \citep{faces}. The dataset consists of greyscale images of faces that were taken between 1992 and 1994 at AT\&T laboratories in Cambridge. There were 40 individuals with 10 images of each individual for a total of 400 $64\times 64$ images. The images were taken with varied lighting, expressions (eyes open/closed, smile/frown etc.), and glasses or no glasses. We fit the model for 15 to 30 column and row factors, and for $G=1,\ldots, 9$ components. The BIC chooses three components with 23 column factors and 26 row factors. The estimated mixing proportions are $\pi_1=0.22,\pi_2=0.49,\pi_3=0.29$.  In Figure \ref{fig:Face}, we show a heatmap of the estimated location parameters for each component. The heatmap for component 3 arguably shows the clearest image and appears to display the glasses feature. 

Upon looking at individual faces classified to component 3 (Figure~\ref{fig:Face}), all the faces have glasses. Moreover, all faces with glasses are classified to component 3 with the exception of two which are classified to component~2. The faces with closed eyes are scattered throughout the three different components and are not classified to any one component. Although it is a difficult to determine the main feature that differentiates component 1 from component~2, it is apparent that the eyebrows for the faces classified to component 1 tend to be more prominent and higher above the eyelid. Of course, a semi-supervised approach to these data could be used to detect specific classes, similar to the MNIST analysis (Section~\ref{sec:mnist}). However, the unsupervised analysis here has shown that the MMVBFA approach can be effective at detecting subgroups without training.

\section{Summary}
In this paper, we developed a MMVBFA model for use in clustering and classification of matrix variate data. Two simulations as well as two real data examples were used for illustration. For each of the simulations, the correct number of components and column/row factors were chosen by the BIC for all of the datasets. Perfect classification performance was also obtained in the simulations. In the MNIST digit application, even with a lower level of supervision, we obtained better results than \cite{gallaugher17c}. However, this is probably due to the fact that the MMVBFA model could use the full $28\times 28$ image. In the faces application, the BIC chooses three groups with the third group being defined by the presence of the glasses facial feature. 
The matrix normality of $\fX$ in the MMVBFA model will allow for direct extensions to mixtures of matrix variate $t$ factor analyzers, as well as skewed matrix variate factor analyzers analogous to their multivariate counterparts. 

\end{document}